\documentclass[pdftex,cite,aps,pre,superscriptaddress,amsmath,amsfonts,amssymb,twocolumn]{revtex4}
\renewcommand{\vec}[1]{\mathbf{#1}} 
\usepackage[pdftex]{graphicx}

\newcommand{\figref}[1]{Fig.~\ref{fig:#1}}
\newcommand{\Figref}[1]{Figure~\ref{fig:#1}}
\newcommand{\figreftwo}[2]{Figs.~\ref{fig:#1} and~\ref{fig:#2}}
\newcommand{\Figreftwo}[2]{Figures~\ref{fig:#1} and~\ref{fig:#2}}
\renewcommand{\eqref}[1]{Eq.~\ref{eq:#1}}
\newcommand{\eqreftwo}[2]{Eqs.~\ref{eq:#1} and ~\ref{eq:#2}}
\newcommand{\Eqref}[1]{Equation~\ref{eq:#1}}
\newcommand{\citeasnoun}[1]{Ref.~\onlinecite{#1}}
\newcommand{\secref}[1]{Sec.~\ref{sec:#1}}

\begin{document}

\title{Optical-approximation analysis of sidewall-spacing effects on \\ the force between two squares with parallel sidewalls} 

\author{Saad Zaheer}
\affiliation{Department of Physics, Massachusetts Institute of Technology, Cambridge, MA 02139}
\author{Alejandro W. Rodriguez}
\affiliation{Department of Physics, Massachusetts Institute of Technology, Cambridge, MA 02139}
\author{Steven G. Johnson}
\affiliation{Department of Mathematics, Massachusetts Institute of Technology, Cambridge, MA 02139}
\author{Robert L. Jaffe}
\affiliation{Center for Theoretical Physics and Laboratory for Nuclear Science, Massachusetts Institute of Technology, Cambridge, MA 02139}

\begin{abstract}
Using the ray-optics approximation, we analyze the Casimir force in a
two dimensional domain formed by two metallic blocks adjacent to
parallel metallic sidewalls, which are separated from the blocks by a
finite distance $h$. For $h > 0$, the ray-optics approach is not exact
because diffraction effects are neglected. Nevertheless, we show that
ray optics is able to qualitatively reproduce a surprising effect
recently identified in an exact numerical calculation: the force
between the blocks varies non-monotonically with $h$.  In this sense,
the ray-optics approach captures an essential part of the physics of
multi-body interactions in this system, unlike simpler
pairwise-interaction approximations such as PFA.  Furthermore, by
comparison to the exact numerical results, we are able to quantify the
impact of diffraction on Casimir forces in this geometry.
\end{abstract}

\maketitle 

\section{Introduction}
\label{sec:intro}

\begin{figure}[b]
\begin{center}
\includegraphics[width=0.35\textwidth]{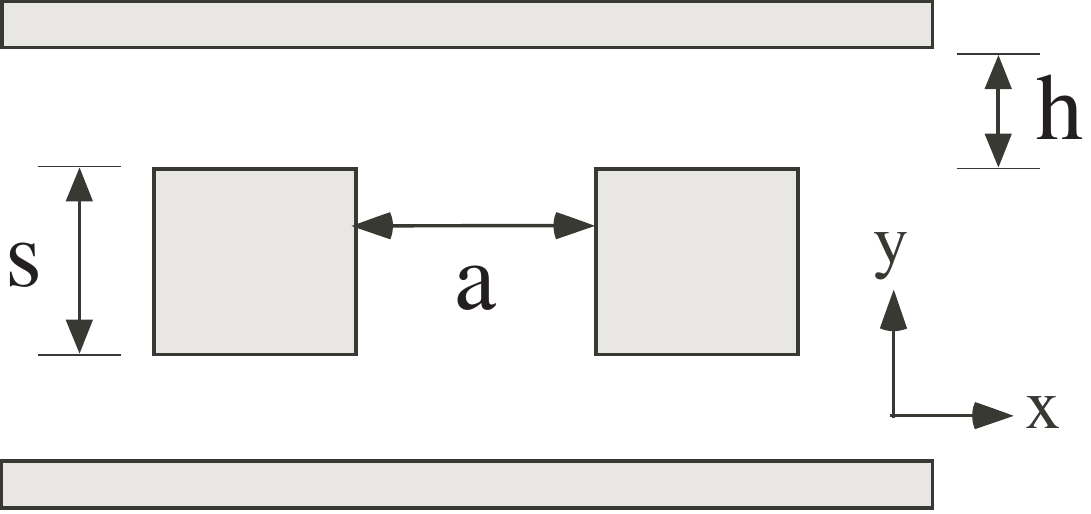}
\caption{Schematic of a two-dimensional geometry: two metal squares $s
  \times s$ separated by a distance $a$, and separated from two
  adjacent metal sidewalls by a a distance $h$.}
\label{fig:geom}
\end{center}
\end{figure}

Casimir forces, which arise from quantum vacuum fluctuations between
uncharged surfaces~\cite{casimir, milonni,Lifshitz80, Plunien86,
Mostepanenko97}, have attracted increasing interest in recent years
due to rapidly improving experimental capabilities for nanoscale
structures~\cite{Lamoreaux97,Onofrio06,Capasso07:review}. At the same
time, theoretical efforts to predict Casimir forces for geometries
very unlike the standard case of parallel plates have begun to yield
fruit, with several promising ``exact'' (arbitrary accuracy) numerical
methods having been demonstrated for a few strong-curvature
structures~\cite{emig06, gies06:edge, Rodriguez07:PRL, Emig07}. In
this paper, we explore the ability of a simple approximate method, the
ray-optics technique~\cite{Jaffe04,Jaffe04:preprint}, to bridge the
gap between analytical calculations for simple geometries and
brute-force numerics for complex structures.  Unlike
pairwise-interaction approximations such as the proximity-force
approximation (PFA)~\cite{bordag01}, ray optics can capture multi-body
interactions and thus has the potential to predict phenomena that
simpler techniques cannot.  In particular, we show that the ray-optics
approach can qualitatively predict a recently discovered
~\cite{Rodriguez07:PRL,Rodriguez07:PRA} non-monotonic effect of
sidewall separation on the force between two squares adjacent to
parallel walls, as depicted in \figref{geom}.  For a sidewall
separation $h=0$, this is known as a ``Casimir piston,'' and in that
case has been has been solved
exactly~\cite{Hertzberg07:notes,Cavalcanti04}. While the ray optics
approach is exact in this structure only for the piston case, its
ability to capture the essential qualitative features for $h > 0$
suggests a wider utility as a tool to rapidly evaluate different
geometries in order to seek interesting force phenomena.  Furthermore,
by comparison to an exact brute-force numerical
method~\cite{Rodriguez07:PRA}, we can evaluate the precise effect of
diffraction (which is neglected by ray optics) on the Casimir force in
this geometry.

The ray optics approximation expresses the Casimir force in terms of a
sum of contributions from all possible classical ray paths (loops)
with the same starting and ending point~\cite{Jaffe04}.  While it is
strictly valid only in the limit of low surface curvature, since it
neglects diffraction effects, the rays include multiple-body
interactions because there exist loops that bounce off multiple
objects.  In contrast, most other low-curvature approximations, such
as PFA~\cite{bordag01} or other perturbative
expansions~\cite{Sedmik06,Schaden98}, are essentially
pairwise-interaction laws, and can therefore miss interesting physics
that occurs when multiple bodies are brought together.  One example
occurs in the structure depicted in \figref{geom}, where there is a
force between two square ($s \times s$) metallic blocks separated by a
distance $a$ that is affected by the presence of two infinite parallel
metallic sidewalls, separated from the blocks by a distance $h$.  For
perfect metals in the $h=0$ limit, this geometry was solved
analytically in both two dimensions~\cite{Cavalcanti04} for Dirichlet
boundary conditions and in three dimensions~\cite{Hertzberg07,
Marachevsky07} for electromagnetic fields. (By ``two dimensions,'' we
mean three-dimensional electromagnetism restricted to $z$-invariant
fields; equivalently, a combination of scalar waves with Dirichlet and Neumann boundary conditions, corresponding to the two polarizations.) For $h > 0$, this geometry was recently solved by an exact
computational method (that is, with no uncontrolled approximations)
based on numerical evaluation of the electromagnetic stress
tensor~\cite{Rodriguez07:PRL,Rodriguez07:PRA}.  In this case, an
unusual effect was observed: as $h$ is increased from $0$, the force
between the blocks varies \emph{non-monotonically} with $h$. The
attractive force between the squares actually decreases with $h$ down to
some minimum and then increases toward the asymptotic limit of two
isolated squares $h\rightarrow \infty$. In the numerical solution,
this non-monotonic effect arose as a competition between the TM
polarization (electric field in the $z$ direction, with Dirichlet
boundary conditions) and the TE polarization (magnetic field in the
$z$ direction, with Neumann boundary conditions), which have opposite
dependence on $h$.  As explained below, it is unclear how this
non-monotonic effect could arise in PFA or similar methods---even if
sidewall effects are included by restricting the pairwise force due to
line-of-sight interactions, it seems that the effect of the sidewall
must always decrease monotonically with $h$.  When we analyze this
structure in the ray-optics approximation, however, we find that a
similar competition between the loops with an even and odd number of
reflections again gives rise to a non-monotonic $h$ dependence.

Below, we first give a general outline of the ray-optics approach,
explain why pairwise approximations such as PFA must fail
qualitatively in this geometry, and then present our results for the
structure of \figref{geom} in two dimensions.  This is followed by a
detailed description of the ray-optics analysis for this structure,
which involves a combination of analytical results for certain
(even-reflection) paths and a numerical summation for other paths.

\section{Ray-optics Casimir forces}
\label{sec:ray}

Following the framework of \citeasnoun{Jaffe04}, we express the
two-dimensional Casimir energy via the ray-optics approximation. The
ray-optics approach recasts the Casimir energy as the trace of the
(scalar) electromagnetic Green's function $G(\vec{x},\vec{x}') =
[\nabla^2 - \omega^2\varepsilon(\vec{x},\omega)]^{-1}
\delta(\vec{x}-\vec{x}')$, which is in turn expressed as a sum over
contributions from classical ``optical'' paths via saddle-point
integration of the corresponding path-integral (this is also referred
to as the ``classical optical approximation''). The optical paths
follow straight lines and are labeled by the number of specular
reflections from the surfaces of the conducting objects. In
particular, the Casimir energy between flat surfaces for Dirichlet
($\eta=-1$) or Neumann ($\eta=1$) boundary conditions is given
approximately by~\cite{Jaffe04}:
\begin{equation}
  \mathcal{E}_r = - \frac{\hbar c}{4\pi} \sum_{r} \eta^r \int_{D_r}
  \frac{d^2x}{\ell_r(\vec{x})^3}
  \label{eq:energy-general}
\end{equation}
Here, the length of a closed geometric path starting and ending at a
point $\vec{x}$ is denoted by $\ell_r(\vec{x})$. $D_r$ is the set of
points that contribute to a closed optical path reflecting $r$ times
from the conducting surfaces. The Casimir energy above is thus the
integral over the whole domain of such points. This problem reduces to
computing a term-by-term contribution from each possible closed path,
as determined by the specific geometrical features of the system under
consideration. Because the Neumann (TE) and Dirichlet (TM) boundary
conditions are given by the sum and difference of the even and odd
paths (paths with even/odd numbers $r$ of reflections), respectively,
it is convenient to compute the contribution of even and odd paths
separately.  The Casimir force is then obtained by the derivative of
the energy with respect to the object separation $a$.

\Eqref{energy-general} is exact for objects with zero curvature (flat
surfaces).  In the presence of curved surfaces (or sharp corners,
which in general have measure-zero contribution to the set of ray
paths), however, the energy will include additional diffractive
effects that are not taken into account by \eqref{energy-general}.
One can include low-order corrections for small
curvature~\cite{Jaffe04:preprint}, but this is obviously not
applicable to the case of sharp corners.  There is one special
exception, the $h=0$ ``piston'': in this case, the sum over optical
paths reduces to the method of images, which is exact for the interior
of rectangular structures.  These limitations are to be expected,
however, since the optical theorem is a stationary-phase
approximation.

\section{Pairwise-interaction approximations}
\label{sec:pairwise}

There are various pairwise-interaction force laws that have been
proposed as approximate methods to compute Casimir forces in arbitrary
geometries.  The most well-known of these is the proximity-force
approximation (PFA), which treats the force between two bodies as a
pairwise sum of ``parallel-plate'' contributions~\cite{bordag01}.  PFA
is exact for parallel plates, and may have low-order corrections
for small curvature~\cite{Bordag06}, but is an uncontrolled
approximation for strong curvature where it can sometimes give
qualitatively incorrect
results~\cite{emig01,genet03,emig03_1,gies06:PFA,maianeto05,Rodriguez07:PRL}.
Another pairwise interaction is the Casimir-Polder $1/r^7$ potential,
valid in the limit of dilute media, which has recently been proposed
as a simple (uncontrolled) approximation for arbitrary geometries by
renormalizing it for the parallel-plate
case~\cite{Casimir48:polder,Sedmik06}.  In this section, we briefly
argue why no such pairwise-interaction approximation can give rise to
the non-monotonic dependence on $h$ that we observe in the structure
of \figref{geom}.

Of course, if one considers the pairwise interaction as a true
two-body force, for each pair of bodies in isolation, then the
sidewalls in the structure can have no effect whatsoever: the force from
one sidewall on one square will be exactly vertical (and cancelled by
the force from the other sidewall).  However, a ``lateral force'' from
the sidewalls, or equivalently an $h$-dependent change in the
attractive force between the two squares, can be obtained by
restricting the pairwise interactions to ``line of sight'' forces.
For example, when considering the force on one vertical edge of a
square from one of the sidewalls, one would include contributions only
from the portion of the sidewall that is ``visible'' from that edge
(connected in a straight line from a point on the edge to the point on
the sidewall without passing through either
square)~\footnote{Such a restriction is at best \emph{ad hoc},
since the points within each body are not ignored even though they are
``blocked'' by the covering material. Alternatively, one could
formulate the two-body interaction as a force between surface
elements exposed to one another, but then there are ambiguities about
how to treat the surface orientation.}.  For a fixed $h$, the
line-of-sight force on the left edge of a square will be different
from the force on the right edge, since one edge will have a portion
of the sidewall blocked by the other square, and hence there will be
an $h$-dependence of the horizontal attractive force.

In particular, since the outside edges of the squares ``see'' (and are
attracted to) a greater portion of the sidewalls than the inside
edges, the net force from the sidewalls will \emph{always reduce} the
attractive force between the squares.  Already, this contradicts the
exact numerical calculations, in which both the Neumann force and the
total force are \emph{greater} at $h=0$ than for $h\rightarrow\infty$.

Moreover, the effect of the sidewalls in a pairwise approximation must
always \emph{decrease} with $h$, again contradicting our results and
making non-monotonic effects impossible.  As $h$ increases, two things
happen: first, the inner edges of the square ``see'' a larger portion
of the sidewalls, with area proportional to $h$; second, the distance
from the sidewalls to the squares increases proportional to $h$.  The
latter contribution must always dominate, however, because any
pairwise force must decrease at least as fast as $1/h^3$ in two
dimensions in order to reproduce the parallel-plate result.
Therefore, the sidewall contribution must decrease monotonically at
least as fast as $1/h^2$ in any pairwise-interaction approximation.

Unlike pairwise-interaction approximations, we show below that the
ray-optics approximation correctly reproduces both qualitative
behaviors: the total force is larger for $h=0$ than for
$h\rightarrow\infty$, and the total force is non-monotonic in $h$.

\section{Results}
\label{sec:results}

Here, we present the results of our calculations for the general $h
\geq 0$ structure shown in \figref{geom}, and compare with the numerical
results from the stress-tensor method~\cite{Rodriguez07:PRL,
Rodriguez07:PRA}. It turns out that the ray-optics technique indeed
captures the non-monotonic dependence of the force with $h$, although
of course the quantitative predictions differ from the exact
calculations. By definition of ray optics, these quantitative corrections
can be attributed to diffraction from the corners. Because we
wish to emphasize the results of the ray-optics approach, rather than
the details of the calculation of the different loop-lengths $\ell_r$
in \eqref{energy-general}, we defer those calculational details until
\secref{details} and here discuss the results.

\begin{figure}[t]
\begin{center}
\includegraphics[width=0.48\textwidth]{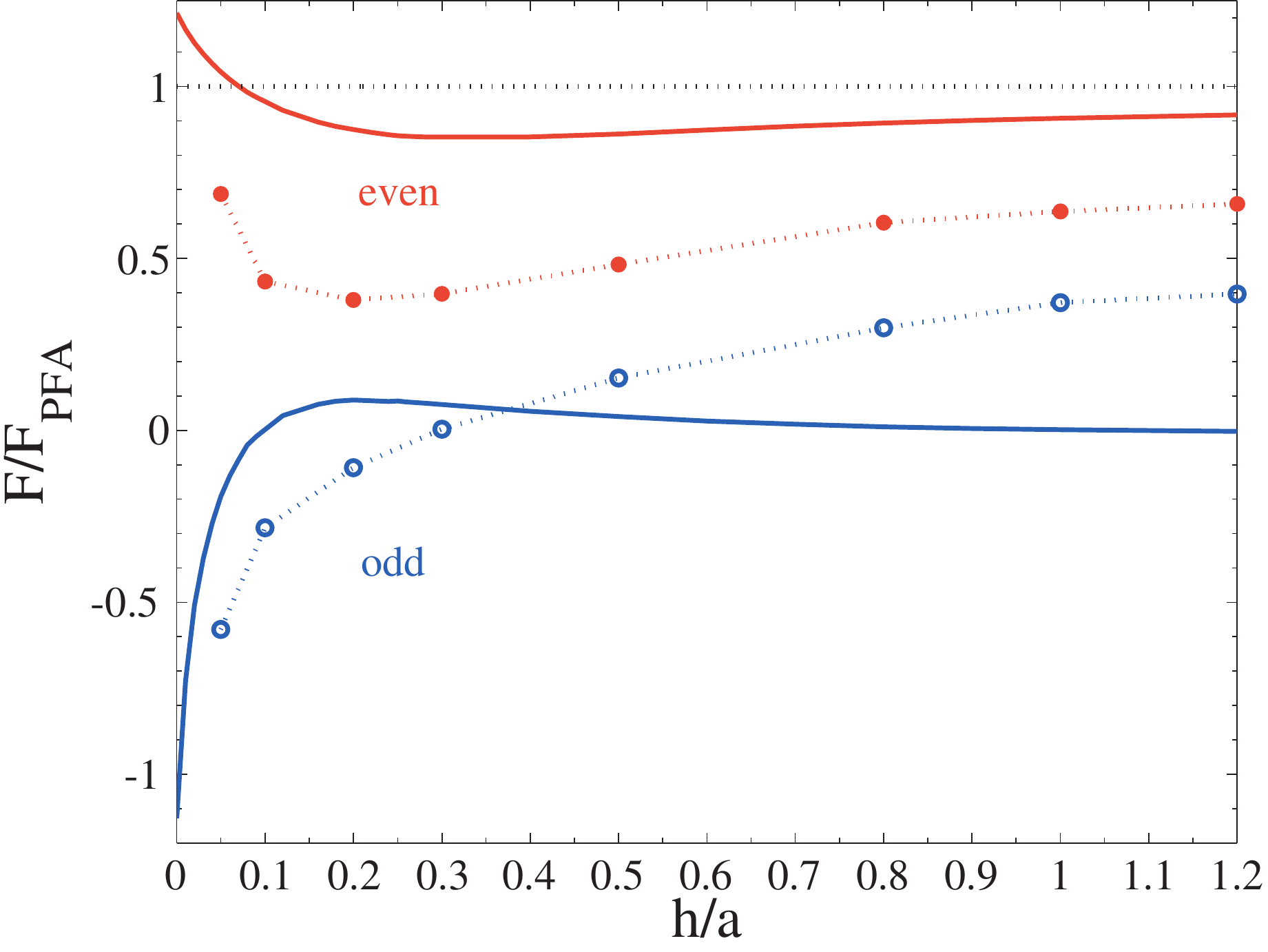}
\caption{Casimir even (red) and odd (blue) forces vs. sidewall
  separation $h$ normalized by the the PFA force
  $F_{\text{PFA}}=-\hbar c \zeta(3)s/8\pi a^3$ (dashed black),
  computed using the ray-optics (solid) and stress-tensor (dashed)
  methods. Note that the ray-optics results become exact as
  $h\rightarrow 0$.}
\label{fig:force1}
\end{center}
\end{figure}

\begin{figure}[t]
\begin{center}
\includegraphics[width=0.48\textwidth]{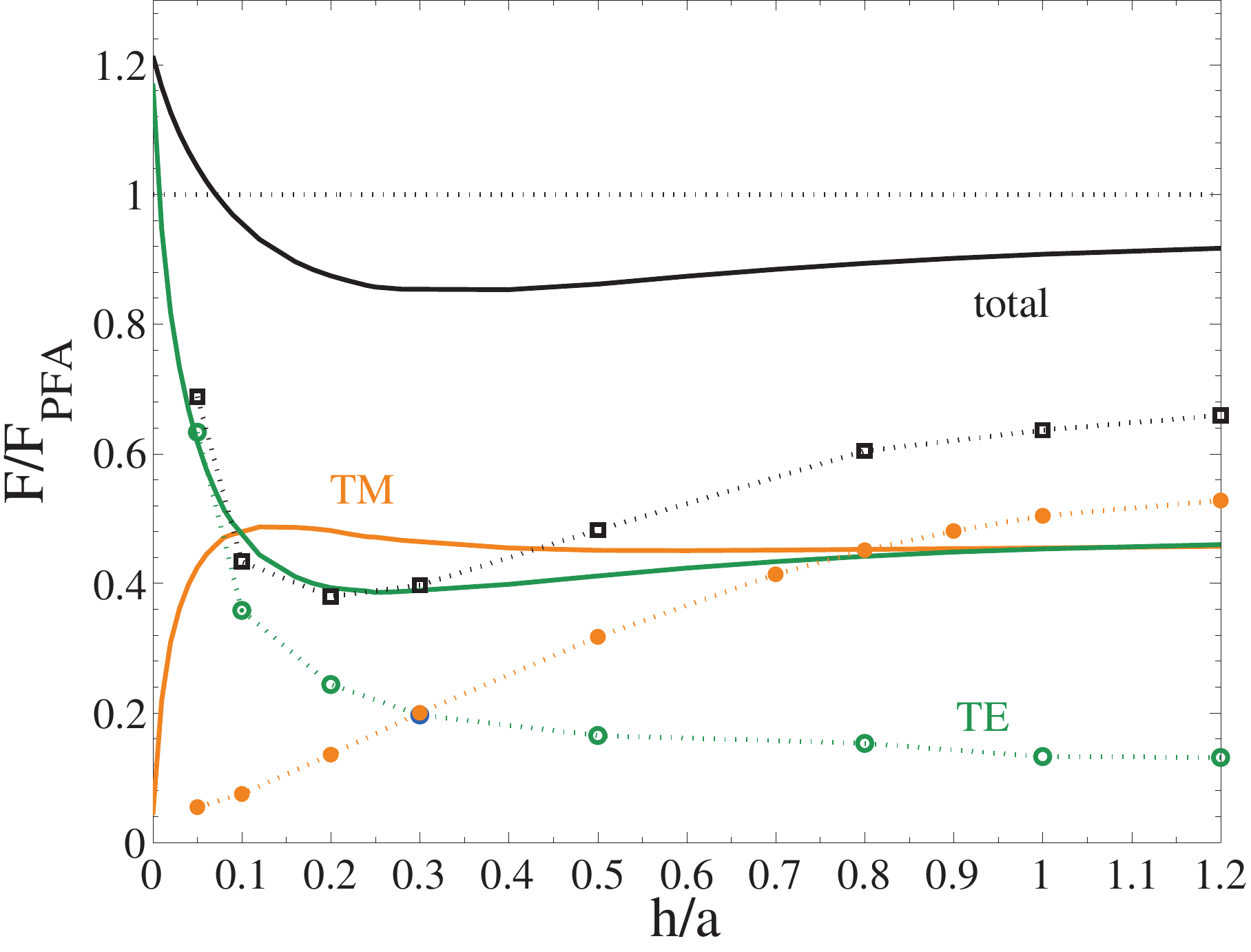}
\caption{Casimir force vs. sidewall separation $h$ normalized by the
  the PFA force $F_{\text{PFA}}=-\hbar c \zeta(3)s/8\pi a^3$, computed
  using the ray-optics (solid) and stress-tensor (dashed) methods. The
  Neumann (green), Dirichlet (orange) and total (black) forces are all normalized
  by the total Neumann+Dirichlet PFA force.}
\label{fig:force2}
\end{center}
\end{figure}

\Figreftwo{force1}{force2} show two different plots of the force
vs. distance from the metal sidewalls $h$, computed via both
\eqref{energy-general} (solids) and the numerical stress-tensor method
(dashed). All results are normalized by the PFA force between isolated
squares (see captions), which are independent of $h$. The bottom panel
shows the contributions from Neumann boundaries (TE polarization) and
Dirichlet boundaries (TM polarization), along with the total
Neumann+Dirichlet force.  Recall that, in the ray optics
approximation, the Neumann/Dirichlet forces are given in terms of the
even- and odd-path contributions by
$(\mathrm{even}\pm\mathrm{odd})/2$, respectively, and thus the total
Neumann+Dirichlet force is equal to the contribution of the even paths
alone.  Because the even/odd decomposition is more natural, in the ray
optics approximation, than Neumann/Dirichlet, the top panel shows the
even/odd contributions from the same calculations.  (Although the
stress-tensor calculation does not decompose naturally into even and
odd ``reflection'' contributions, here we simply define the even/odd
components as $(\mathrm{Neumann}\pm\mathrm{Dirichlet})/2$,
respectively.)

As $h$ goes to zero, the ray-optics results become exact. The
numerical computation of the stress-tensor force becomes difficult for small
$h$ due to our implementation's uniform grid, but nevertheless the
linear extrapolation of the numerical calculations to $h=0$ agree with
ray optics to within a few percent. For
$h >0$, the total force for both the ray-optics and stress-tensor
results displays a minimum in the range $h\sim 0.2$--$0.3$. In
particular, the extrema lie at $h\approx 0.3$ and $h\approx 0.25$,
respectively.  Not only is this striking non-monotonic behavior
captured by the ray-optics approximation, but the agreement in the
location of the extremum is also excellent.

Thus far, \figreftwo{force1}{force2} reveal two significant
differences between the ray-optics and stress-tensor results. First,
the forces when $h$ is not small differ quantitatively, by about 30\%
as $h\rightarrow \infty$.  Since the ray-optics approximation is
essentially obtained by dropping terms due to diffraction (from curved
surfaces and corners), we can attribute this quantitative difference
to the diffractive contribution to the Casimir force from the finite
size.  In the large-$h$ limit, where the sidewalls become irrelevant
and the ray-optics result approaches PFA, the differences compared
to the exact solution are sometimes called \emph{edge
effects}~\cite{Rodriguez07:PRL,Rodriguez07:PRA,gies06:edge}. Second,
although the exact and ray-optics results match in the $h=0$ limit as
discussed above, the functional forms for small $h$ are quite
different, at least for Dirichlet boundaries.  In the ray optics
expressions, the odd contributions have a logarithmic singularity at
$h=0$, which lead to corresponding singularities in the Neumann and Dirichlet
forces.  However, in the exact stress-tensor calculation, only the Neumann
force seems to display a sharp upturn in slope as $h=0$ is approached
(although it is impossible to tell whether it is truly singular); the
stress-tensor Dirichlet force seems to be approaching a constant slope (which
is why we were able to linearly extrapolate it to $h=0$ with good
accuracy). 

Another qualitative difference appears if we look at the even and odd
contributions in \figref{force1}: whereas ray optics and the
stress-tensor method give a similar non-monotonic shape for the even
force, the odd forces are quite different.  (The stress-tensor odd
force is monotonic while the ray-optics odd force is not, while the
latter goes to zero for large $h$ and the former does not.)  Again, we
attribute this to a greater sensitivity to diffraction effects, this
time for the odd forces compared to the even forces.  As will be
argued in \secref{even}, the domain of integration and the length of
even ray-optics paths have a weaker dependence on the corners of the
squares than the odd paths, and thus should be less sensitive to
corner-based diffractive effects.  Fortunately, the total force
depends solely on the even path contributions, which helps to explain
why ray-optics ultimately does effectively capture the non-monotonic
behavior and the location of the extremum.

\begin{figure}[htb]
\includegraphics[width=0.45\textwidth]{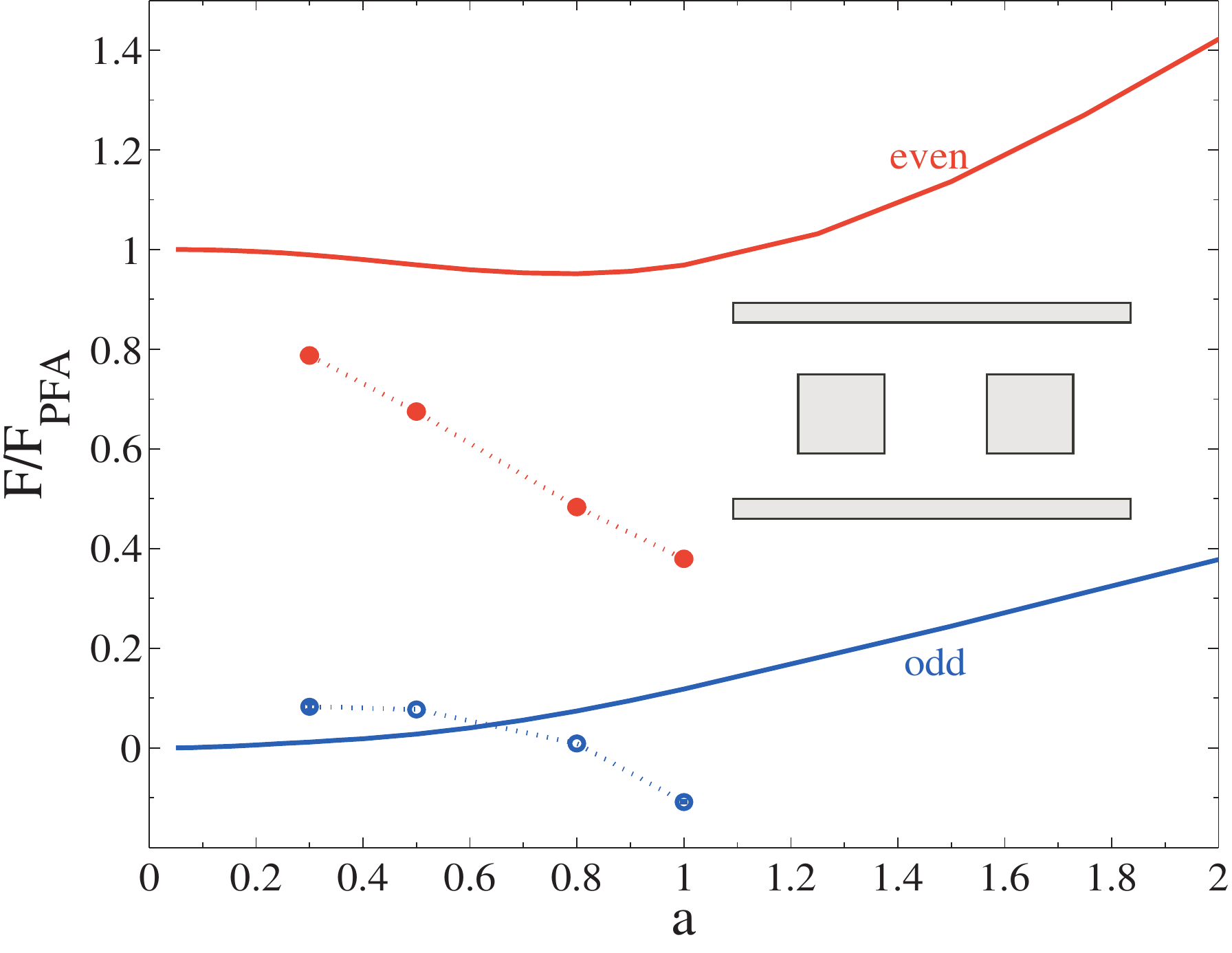}
\includegraphics[width=0.45\textwidth]{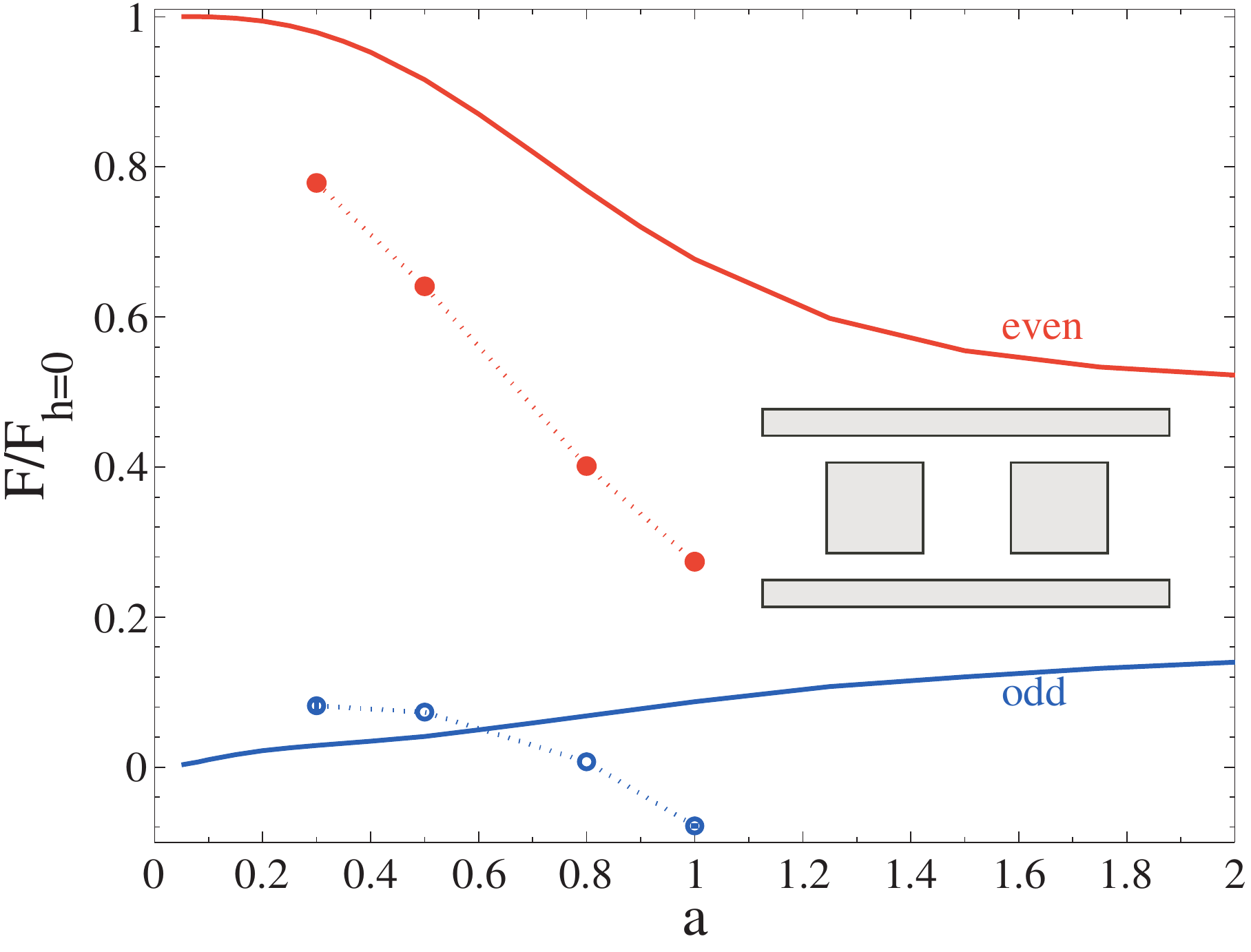}
\caption{Casimir force vs. separation between squares $a$ with
  constant sidewall separation $h=0.25$, normalized by the
  $F_{\text{PFA}}=-\hbar c \zeta(3)s/8\pi a^3$ (top) and the
  corresponding $h=0$ force (bottom). The force is computed for both
  even (red) and odd (blue) contributions, separately. \emph{Inset:}
  schematic of geometry consisting of two isolated squares with
  adjacent sidewalls.}
\label{fig:force-a2}
\end{figure}

Having explored the $h$-dependence of the force using both ray-optics
and numerical stress-tensor methods, we now turn to \figref{force-a2}
to study the behavior of the force as a function of the square
separation $a$. \Figref{force-a2} shows the Casimir force vs. square
separation $a$ at constant $h/s=0.25$, normalized by the PFA force
between isolated squares (top panel) or by the exact force at $h=0$
(bottom panel).  Note that in both cases the normalization is
$a$-dependent, unlike in \figreftwo{force1}{force2}, and any
non-monotonicity in \figref{force-a2} is only an artifact of this
normalization.  The normalization by the PFA force allows us to gauge
both the sidewall/edge effects (which disappear for $a\rightarrow 0$)
and whether there is a difference in scaling from PFA's $1/a^3$
dependence.  For the bottom panel, we normalize against the exact
$h=0$ force, which tells us whether the finite sidewall separation
makes a difference for the large-$a$ scaling.  In both cases, we show
a few points of the stress-tensor calculation (which became very
expensive for small or large $a$), to get a sense of the accuracy of
the ray-optics method at different $a$.

We should expect that as $a\rightarrow 0$, both the PFA and the
ray-optics solution should approach the exact solution, because the
sidewall contribution becomes negligible.  This agreement, as compared
to the extrapolated numerical stress-tensor results, can be observed
in \figref{force-a2}(top).  In contrast, for the large-$a$ limit the
ray-optics force appears to decay as $1/a^2$ instead of $1/a^3$ for
PFA, leading to the apparent linear growth in the top panel of
\figref{force-a2}.  If we compare to the $h=0$ dependence in
\figref{force-a2}(bottom), it appears to be asymptoting to a constant
for large $a$, indicating that the power laws for $h=0$ and $h > 0$
may be identical.  However, even if we had more data it would be
difficult to distinguish the presence of, for example, logarithmic
factors in this dependence.  For the $h=0$ case, we have analytical
results for even- and odd-path forces in \secref{hzero}: from the
analytical expressions, the odd-path $h=0$ force clearly goes as
$1/a^2$, and the even-path $h=0$ force also turns out to have the same
$1/a^2$ dependence~\footnote{The asymptotic behavior of the Epstein
Zeta function in the even force (\secref{hzero}) seems little studied,
but it is straightforward to show that it goes as $1/a^2$ for large
$a$.}.  The exact stress-tensor computation appears to be quite
different from both the PFA and the ray-optics force as a function of
$a$, but we were not able to go to large enough computational cells to
estimate the asymptotic power law.  It is striking that $h/s$ as small
as $0.25$ is already large enough to yield substantial diffraction
effects in the force.

\section{Details of the ray-optics computation}
\label{sec:details}

In this section, we describe the computation of the ray-optics
approximation for the $h>0$ squares+sidewalls structure, according to
\eqref{energy-general}. This involves systematically identifying all
of the possible closed ray loops, integrating them for a given number
of reflections over the spatial domain, and then summing over the
number of reflections.  It turns out that the contributions of any
even reflection order can be integrated analytically as shown below,
although the final summation over reflection order is still numerical.
On the other hand, the integrals from the odd reflection orders become
increasingly difficult as the reflection order is increased, and so we
resorted to numerical integration for odd orders $r > 5$.

\subsection{Even paths}
\label{sec:even}

\begin{figure}[ht]
\includegraphics[width=0.43\textwidth]{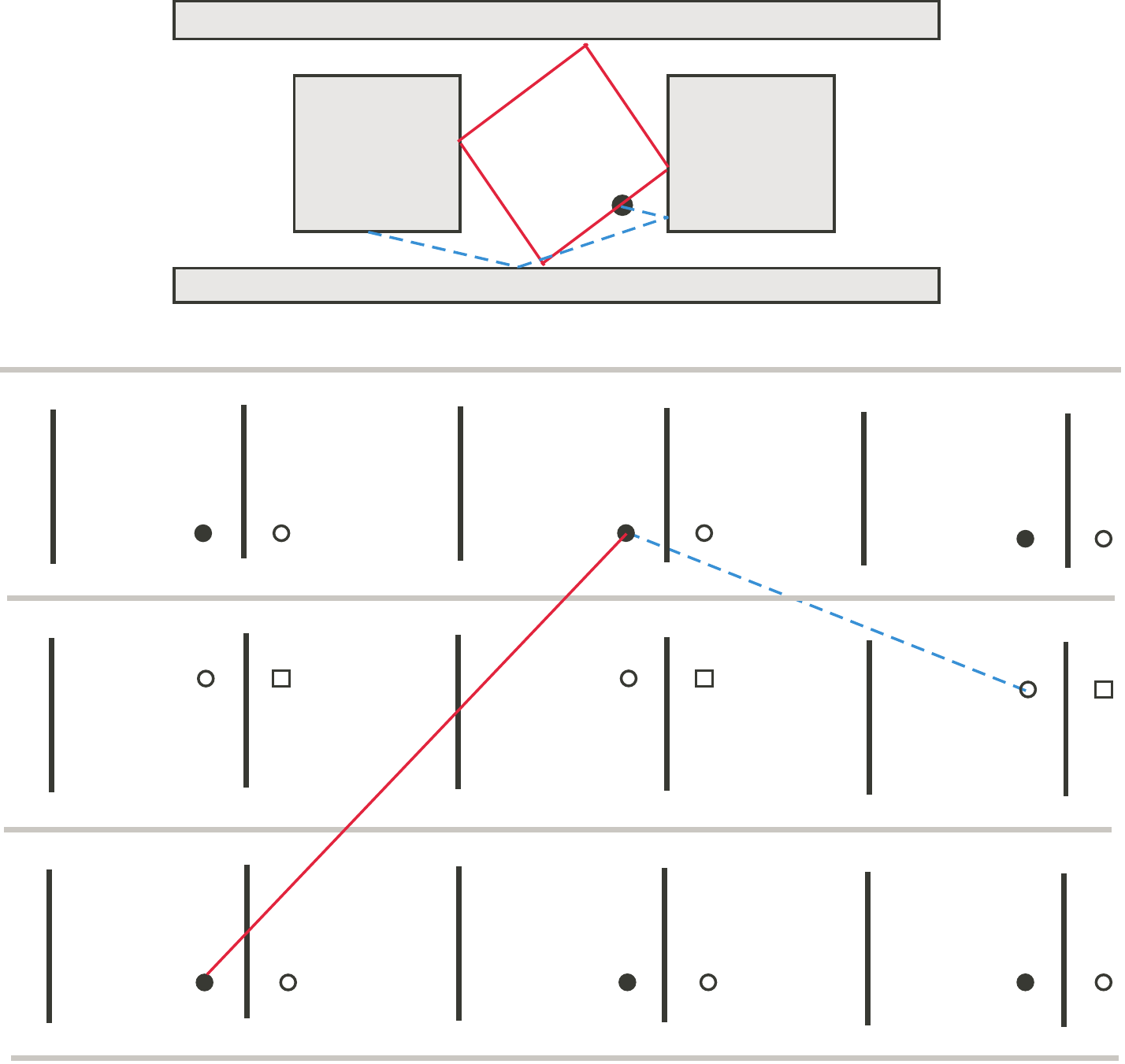}
\caption{Schematic of general 2d squares+sidewalls lattice. Lines extending from
  solid circles unto solid circles represent even reflection
  paths. Lines extending from open solid circles unto open circles
  represent odd reflection paths. (Here, $h/a\approx 0.2$ and
  $s/a\approx 1$.) A possible even (red) and forbidden odd (blue) path
  is shown.}
\label{fig:lattice}
\end{figure}

Because we are dealing with perfect metals, and because the geometry
has reflection-symmetry about the $x$ and $y$ axes, it is helpful to
represent the optical paths using an infinite periodic lattice, shown
in \figref{lattice}, similar to the construction in
\citeasnoun{Hertzberg07}. The reason for this construction is that,
because of the equal-angle law for specular reflections in geometric
optics, a reflected ray is equivalent to a linearly extended ray in a
mirror-image structure.  This allows us to visualize and count the set
of possible closed paths in a straightforward fashion. Specifically, a
closed path which starts at a point $\vec{x}$ and ends on itself is
fully determined by the set of lines that start at $\vec{x}$ and end
in the corresponding set of image points on the extended lattice.  The
unit cell of this periodic construction is just two vertical black
lines (of length $s$ and separation $a$) that represent the parallel
walls of the two squares. These are repeated with a horizontal period $a$ and a vertical period $s+2h$.  Any path that passes through the gap between
one of these lines and the horizontal sidewalls escapes from between
the two squares and therefore is not counted among the closed loops.

In order to construct all of the closed loops that originate at a
given point $\vec{x}$ in the unit cell, we proceed as follows.  First,
we construct the mirror reflections of this point through the vertical
lines (the boundaries of the squares) and the horizontal lines (the
sidewalls), corresponding to reflections from these metallic walls.
This gives us a set of points in the nearest-neighbor cells.  Then, we
construct the reflections of the nearest-neighbor points through their
sidewalls, and so on, corresponding to reflections of higher and
higher order.  A closed loop is simply a line segment from the
original $\vec{x}$ to one of the reflected points, as long as it does
not escape through one of the gaps between the squares and sidewalls
as explained above.  \Figref{lattice} was generated from a unit cell
with $h/a\approx 0.2$ and $s/a\approx 1$, and shows both even-reflection
(solid red) and odd-reflection (dashed blue) paths, where in this case
the odd path shown escapes and therefore would not be counted.  In the
figure, the points are labelled according to the number of reflections
that generate them from the original point: solid circles when the
numbers of horizontal and vertical reflections are both even, open
squares when the numbers are both odd, and open circles otherwise.  It
follows that lines connecting solid circles to solid circles are even
paths, and lines connecting solid to open circles are odd paths; the
rays connecting solid circles and squares always escape and therefore
do not contribute..

To compute the Casimir energy from these paths, the key quantity in
\eqref{energy-general} is the length $\ell$ of the path.  Let us label
each unit cell by $(n,m)$ according to its horizontal ($n$) and
vertical ($m$) offset from the cell $(0,0)$ where the original point
$\vec{x}$ resides.  For an even path, $\vec{x}$ must be connected to
an even-indexed image $(2n,2m)$, for which the length of the path is:
\begin{equation}
  \ell_{n,m} = \sqrt{(2na)^2+\left[2m\left(2h+s\right)\right]^2}
\label{eq:ellnm}
\end{equation}
and the angle of the path, determined by $m/n$, is:
\begin{equation}
  \tan \theta_{n,m} = \frac{m}{n}\frac{2h+s}{a} .
\end{equation}
The only things left to figure out are the domain of integration of
$\vec{x}$ and the allowed $(n,m)$ for non-escaping paths. If $h=0$, it
is obvious that our expression reduces to the expression of
\citeasnoun{Hertzberg07}, since both the whole spatial domain within
the unit cell and \emph{all} $(n,m)$ are allowed. However, when $h >
0$, each $(n,m)$ will be non-escaping only for $\vec{x}$ in a subset
of the unit cell.  To determine these subsets, the domains of
integration, we take advantage of the closed-loop nature of the paths
to cast \eqref{energy-general} in a different light.  For a given
$(n,m)$, instead of integrating over $x$ and $y$, it is convenient to
change variables to integrate over $y$ and a coordinate $\xi = x /
\cos\theta_{n,m}$ measuring displacement along the path (along the
$\theta_{n,m}$ direction).  It might seem that one should integrate
$\xi$ along the whole line from $(0,0)$ to $(n,m)$, but this may
involve counting the same point $\vec{x}$ in the unit cell multiple
times.  Instead, to avoid over-counting loops that wrap around on
themselves, one instead integrates from $(0,0)$ to $(\tilde{n},
\tilde{m})$ where $\tilde{n}=n / \gcd(n,m)$ and
$\tilde{m}=m/\gcd(n,m)$ are reduced to lowest terms.  One then obtains
the following equation for the energy:
\begin{equation}
  \mathcal{E}_\mathrm{even} = -\frac{\hbar c}{4\pi} \sum_{n,m} \int dy \int_{\xi(0,0)}^{\xi(\tilde{n},\tilde{m})}
  d\xi \cos \theta'_{\tilde{n},\tilde{m}}
  \frac{1}{\ell^{3}_{n,m}}
\label{eq:energy-angle}
\end{equation}

Although not so obvious from looking at \eqref{energy-angle}, the
integral in the $y$ direction simply counts the number of
$(\tilde{n},\tilde{m})$ paths that exist in the unit cell. Because the
length and angle of such paths are independent of $y$, we can
integrate over $d\xi$ to obtain:
\begin{equation}
  \mathcal{E}_\mathrm{even} = -\frac{\hbar c}{4\pi} \sum_{n,m} \int dy
  \frac{2\tilde{n}a}{\ell^3_{n,m}}
\end{equation}
where we used $\int d\xi = \ell_{\tilde{n},\tilde{m}}$ and $\cos
\theta_{\tilde{n},\tilde{m}} =
2\tilde{n}a/\ell_{\tilde{n},\tilde{m}}$, between which the
$\ell_{\tilde{n},\tilde{m}}$ cancels. All that is left to figure out is
the integral in the $y$ direction.

To carry the integral in the $y$ direction, we must determine the
limits of integration, or equivalently, the range over which we can
displace the path so that it does not escape. We go back to
\figref{lattice} for reference. Again, as outlined above, we extend a
line from a solid circle in the $(0,0)$ cell to another solid circle
in the $(2n,2m)$-th cell. For the path to be allowed, it must
intercept all of the vertical black line segments that lie between the
two points, i.e. at each horizontal reflection from the
squares. Because each interception (horizontal reflection) occurs at
periodic intervals, these end up partitioning the unit cell in the $y$
direction into $\tilde{n}$ sets of length $(2h+s)/\tilde{n}$. Note
that that we divide by $\tilde{n}$, rather than $n$, because as
explained above the topologically distinct paths are uniquely
specified by $(n,m)$ reduced to lowest terms.  From this simple
argument, we obtain that the vertical displacement is
$(2h+s)/\tilde{n} - 2h$, provided that $(2h+s)/\tilde{n} - 2h > 0$. To
help visualize this result, it is best to think of the problem on a
circle. That is, consider a circle of length $2h+s$ and partition it
into $\tilde{n}$ sets of length $(2h+s)/\tilde{n}$, as well as into
two regions of length $s$ and $2h$. If the path is to exist, each of
the $(2h+s)/\tilde{n}$ points on the circle must not intercept the
region marked as belonging to $2h$ (the air gaps between the
squares). The result follows directly by considering the distance that
one can displace the points before any of them intercepts the $2h$
region.

Thus, the final expression for the even path energies is left as a sum
over $n$ and $m$:
\begin{widetext}
\begin{equation}
  \mathcal{E}_\mathrm{even} = -\frac{\hbar c}{4\pi} \sum_{n,m >
    0}\Theta\left(\frac{2h+s}{\tilde{n}}-2h\right)\left[\frac{2h+s}{\tilde{n}}-2h\right]
  \frac{4\tilde{n}a}{\left[(2na)^2+ (2m(2h+s))^2\right]^{3/2}}
\label{eq:energy-even}
\end{equation}
\end{widetext}
An extra factor of 2 was included in the numerator since the
contributions of the $(-n,-m)$ paths are identical to those of the
$(n,m)$ paths by symmetry. In the limit $h=0$, we recover the even
energy expression of \citeasnoun{Hertzberg07:notes}, given in
\secref{hzero}.

Although we are almost done with the even reflection paths, we are
missing a very important contribution to the energy: the PFA terms,
i.e. the $(n,0)$ and $(0,m)$ paths. The PFA energy between two parallel finite
metal regions is a well-known result, and we
include here only $(n,0)$ for completeness:
\begin{eqnarray}
  \mathcal{E}_{\mathrm{PFA}} &=& -\frac{\hbar c s a}{16 \pi}
  \sum_{n>0}\frac{1}{(na)^3} \nonumber \\
  &=& -\frac{\hbar c s \zeta(3)}{16\pi a^2}
\label{eq:}
\end{eqnarray}

The final expression \eqref{energy-even} must still be summed
numerically over $(n,m)$ up to some upper cutoff, but this is an
simple computation and its convergence with the cutoff is discussed in
\secref{num}.

\subsection{Odd paths}
\label{sec:odd}

Unlike even reflections, odd reflection paths are quite tedious to
compute because have less symmetry. For one thing, the length of a
path depends not only on $(n,m)$ (the offset of the unit cells being
connected), but also on the position $\vec{x}$ of the starting point.
More importantly, the identification of the domain of $\vec{x}$ for
non-escaping paths depends in a much more complicated way on $(n,m)$,
making it difficult to write down a single expression that works for
all $(n,m)$.  Therefore, we analytically solved for the odd-path
contributions only up to five-reflection paths, where each order
requires a separate analysis, and treated higher-order paths by a
purely numerical approach.  Below, the analytical solution for the
third-order paths is given, both to illustrate the types of computations
that are involved and also to demonstrate the logarithmic singularity
in the force as $h \rightarrow 0$.

The results of \citeasnoun{Hertzberg07} give an upper limit for the
number of odd paths $r \geq 3$ that exist in this geometry for $h=0$
(\# paths = $2^{(r+3)/2}$).  The same upper limit holds for $h > 0$,
but in this case the number of paths is actually reduced because some
of the $h=0$ paths now escape.  An analytical solution for any
particular order must begin by drawing all paths for $h=0$ and then
perturbing them for $h>0$ to eliminate any impossible paths. At least
for low-order paths, simple geometrical arguments can then determine
the domain of integration.

The coordinate dependence of odd paths arises from the fact that any
path that reflects an odd number of times from the planar surfaces
will be non-periodic: if we extend the path beyond its endpoint (=
starting point), it will not repeat. This greatly complicates the
analysis.  For example, \figref{lattice} shows one such escaping odd
path. It turns out that as $h$ grows, the domain of integration for
odd paths shrinks and becomes harder to visualize. Moreover, if we
vary $h$, we notice that paths for some $(n,m)$, regardless of their
origin $\vec{x}$, always escape.  The problem of determining the
domain of integration and the allowed $(n,m)$ seems rather difficult
and does not seem to have a general closed-form solution.


\subsubsection{Three-reflection paths}

\begin{figure}[ht]
\includegraphics[width=0.33\textwidth]{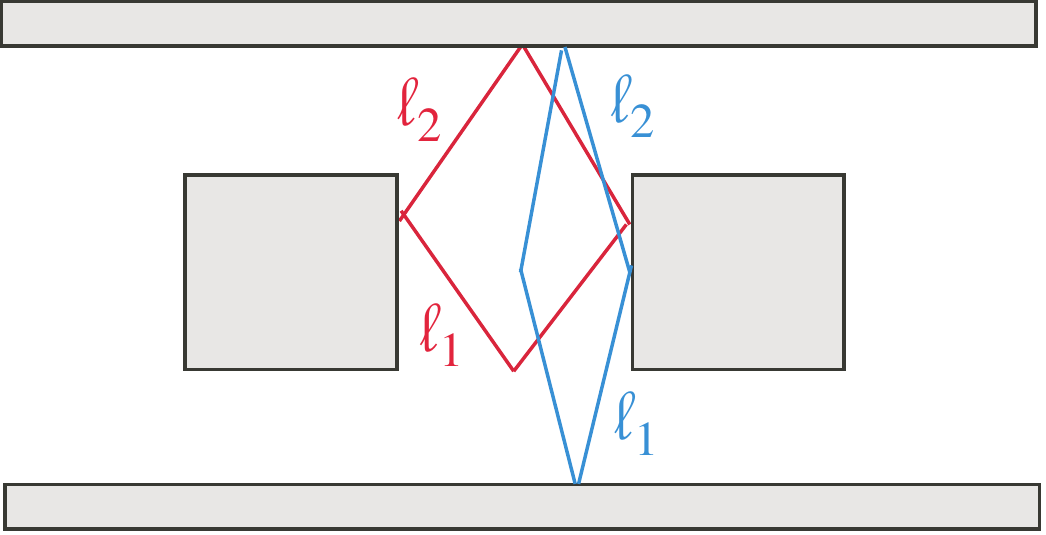}
\caption{Schematic of three-reflection paths. Blue/red represent
  $(1,2)$/$(2,1)$ paths, and the lengths $\ell_{i}$ shown are used the
  calculation of the energy.}
\label{fig:three-r}
\end{figure}

From \figref{lattice} we verify that there are eight possible three
reflection paths, $(\pm 2,\pm 1)$ and $(\pm 1,\pm 2)$
according to the notation of \secref{even}, shown in \figref{three-r}. The only two nonequivalent cases are $(2,1)$ and $(1,2)$, which have different domains of $\vec{x}$ integration.

Inspection of \figref{three-r} and \figref{lattice} yields the length
of these paths $\ell_{2,1} =
2\sqrt{a^2+\left(2h+s-y\right)^2}$. Moreover, we see that $\ell_{2,1}$ is
invariant along the $x$-axis. For this particular path, the maximum
displacement in the vertical direction $y_{\textrm{max}}$ occurs when
$\ell_1 \sin\theta = \ell_2 \sin\theta =h$, while the minimum
$x_{\textrm{min}}$ and maximum $x_{\textrm{max}}$ horizontal
displacements occur when $\ell_2 \sin\theta = h$ and $\ell_1 \sin\theta =
h$, respectively. Therefore, the range of integration is:
\\
\begin{eqnarray}
  0 \leq &y& \leq s
\\ 
\left(\frac{ah}{2h+s-y}\right) \leq &x& \leq
  \frac{a(h+s-y)}{2h+s-y}
,
  \label{eq:nm21}
\end{eqnarray}
\\
giving us the following expression for the energy:
\begin{multline}
    \mathcal{E}_{2,1} = -\frac{\hbar c}{32\pi a^2} \left\{
    \frac{sa}{\sqrt{a^2+\left(2h+s\right)^2}} \right.
    \\ \left. - 2h\log\left[ \left(\frac{2h+s}{2h}\right)
      \frac{1+\sqrt{1+\left(\frac{2h}{a}\right)^2}}{1+\sqrt{1+\left(\frac{2h+s}{a}\right)^2}}
      \right]
    \right\}.
\label{eq:eps_21}
\end{multline}

In order to compute $\mathcal{E}_{1,2}$, we apply similar geometrical
considerations. Once again, from \figref{lattice} we obtain the length
to be $\ell_{1,2} =2\sqrt{\left(a-x\right)^2+
  \left(2h+s\right)^2}$. Similarly, we see that $\ell_{1,2}$ is
invariant along $y$-axis, and approaches a minimum when
$x=a$. Therefore, the range of integration is:
\begin{eqnarray}
h \leq &y& \leq h+s \\ 0 \leq &x& \leq a ,
\end{eqnarray}
giving us the following expression for the energy:
\begin{equation}
\mathcal{E}_{1,2} = -\frac{\hbar c a s}{32 \pi \left(2h+s\right)^2}
\frac{1}{\sqrt{a^2+\left(2h+s\right)^2}} .
  \label{eq:eps_12}
\end{equation}

Adding \eqref{eps_21} to \eqref{eps_12}, and multiplying by four to
account for the different $\pm$ sign possibilities in $(n,m)$ yields
the total three-reflection contribution to the energy.  This
contribution has two types of terms: a polynomial term from
\eqref{eps_12} and the first term of \eqref{eps_21}, and a logarithmic
term from the second term of \eqref{eps_21}.  The polynomial term
remains at $h=0$. The more intriguing component of this result is the
logarithmic term, which falls as $O(h \log h)$ for small $h$,
vanishing completely at $h=0$.  This (and similar terms for
higher-order reflections) is the source of the logarithmic singularity
in slope of the ray-optics odd-path Casimir force at $h=0$ observed in
\figref{force1}.

\subsubsection{Five-reflection paths}
\label{sec:five-reflection}

The analytical solution of the five-reflection contribution is rather
complicated and is not reproduced here.  However, it has a few
interesting features that we summarize here.  First, just as for the
three-reflection paths, the five-reflection contribution has an $O(h
\log h)$ term that contributes to the singularity we observe in the
ray-optics odd-path Casimir force at $h=0$.  Second, one also obtains
$O(\log h)$ terms, which should seem to unphysically diverge as
$h\rightarrow 0$. It turns out, however, that for a path that gives a
$(\log h)$ contribution at small $h$, there exists another path with a
$-(\log h)$ contribution, resulting in exact cancellation of any
divergences.

\subsection{Casimir piston}
\label{sec:hzero}

The $h=0$ limit is the well-known ``Casimir piston'' geometry.  In
this limit, all of the optical paths contribute to the Casimir energy,
making it possible to compute \eqref{energy-general}
analytically. \citeasnoun{Hertzberg07} performs this calculation in
three-dimensions, and a similar unpublished result was obtained in
two-dimensions~\cite{Hertzberg07:notes}.  The two-dimensional geometry
was also solved analytically for Dirichlet boundaries by another
method~\cite{Cavalcanti04}. Here, we reproduce this calculation as a
check on both the even- [\eqref{energy-even}] and odd-path
[\eqreftwo{eps_12}{eps_21}] contributions to the energy, and also
because the Neumann result is useful and previously unpublished.

We begin by computing the $h=0$ even-path contributions. The
expression for the even energy \eqref{energy-even} is much simpler
than for $h> 0$, because the $\Theta$ function disappears and all
that is left is a polynomial function in terms of $n,m$. The even-path
energy, not including the PFA contribution (terms where $n=0$ or
$m=0$, but not both), is given by:
\begin{eqnarray}
\mathcal{E}_{\mathrm{even}} &=& -\frac{\hbar c}{8\pi} \sum_{n,m > 0}
as\left(\left(na\right)^2+ \left(ms\right)^2\right)^{-3/2} \\ &=&
-\frac{\hbar c}{8\pi} as\textrm{Z}_2(a,s;3)
\label{eq:hzero-even}
\end{eqnarray}
we have identified the summation above as the second order
Epstein Zeta function $\textrm{Z}_2$:
\begin{equation} 
\textrm{Z}_2(a,b;p)= \sum_{n,m > 0} \left((na)^2+(mb)^2\right)^{-p/2}
\end{equation}
The same simplification occurs in the case of odd paths, the lengths
of which can be found, again, by inspection of the lattice in
\figref{lattice}. Again, as described in \secref{even}, given a point
on the unit cell $\vec{x}$, one can determine all possible
odd-reflection paths by drawing straight lines from the solid circles
($\vec{x}$) to the open circles in the lattice. 

There are two types of open circles, each of which denote two
different types of paths: those that reflect across the $x$ axis and
have $y$-invariant length, as well as those that reflect across the
$y$ axis and have $x$-invariant length.  Whichever coordinate is the
invariant one gives a constant integral over the unit cell, so the
double integral is reduced to a single integral.  For example,
consider those that reflect across the $x$ axis and have $y$-invariant
length: for these paths, we need to integrate over $x$ in the unit
cell and perform a double sum over $(n,m)$.  However, the double sum
can be reduced to a single sum by eliminating the sum over $n$ in
favor of integrating $x$ along the whole real line instead of just the
unit cell.  Thus, we are left with a single integral and a single
summation.  Similarly paths that reflect across the $y$ axis reduce to
a single integral over $y$ and a single sum over $n$.  As a result of
all this manipulation, the odd integral becomes:
\begin{multline}
\mathcal{E}_{odd} = - \frac{\hbar c s}{32\pi} \sum_{m>0} 
\int_{-\infty}^{\infty} \frac{dx}{[(ms)^2+ x^2]^{3/2}} \\ 
- \frac{\hbar c a}{32\pi} \sum_{n>0} 
\int_{-\infty}^{\infty} \frac{dy}{[(na)^2+ y^2]^{3/2}} .
\label{6.2}
\end{multline}
Again, we restrict the sum to $n,m >0$, because horizontal and
vertical paths are divergent terms that contribute only to the
self-energy of the metallic walls~\cite{Jaffe04:preprint}. Carrying
out the above integrals yields the following expression for the
odd-path energy:
\begin{equation}
\mathcal{E}_{\text{odd}} = - \frac{\hbar c \pi}{48}\left(\frac{1}{s} + \frac{1}{a}\right) .
\label{eq:hzero-odd}
\end{equation} 
The odd-path contribution to the force is therefore $\propto
1/a^2$ for large $a$.
\subsection{Numerics}
\label{sec:num}

To evaluate \eqref{energy-general} numerically, we used an adaptive
two-dimensional quadrature (cubature) algorithm~\cite{Genz83} to
perform the $\vec{x}=(x,y)$ integration for each $(n,m)$.  (For each
$\vec{x}$ and $(n,m)$ it is easy to numerically check whether the path
is allowed, and set the integrand to zero otherwise. Unfortunately,
this makes the integrand discontinuous and greatly reduces the
efficiency of high-order quadrature schemes; an adaptive trapezoidal
rule might have worked just as well.  For very small $h$, this
requires some care because the energy depends on a tiny remainder
between two diverging terms, as we saw in \secref{five-reflection},
but for most $h$ there was no difficulty.

\begin{figure}[htb]
\includegraphics[width=0.47\textwidth]{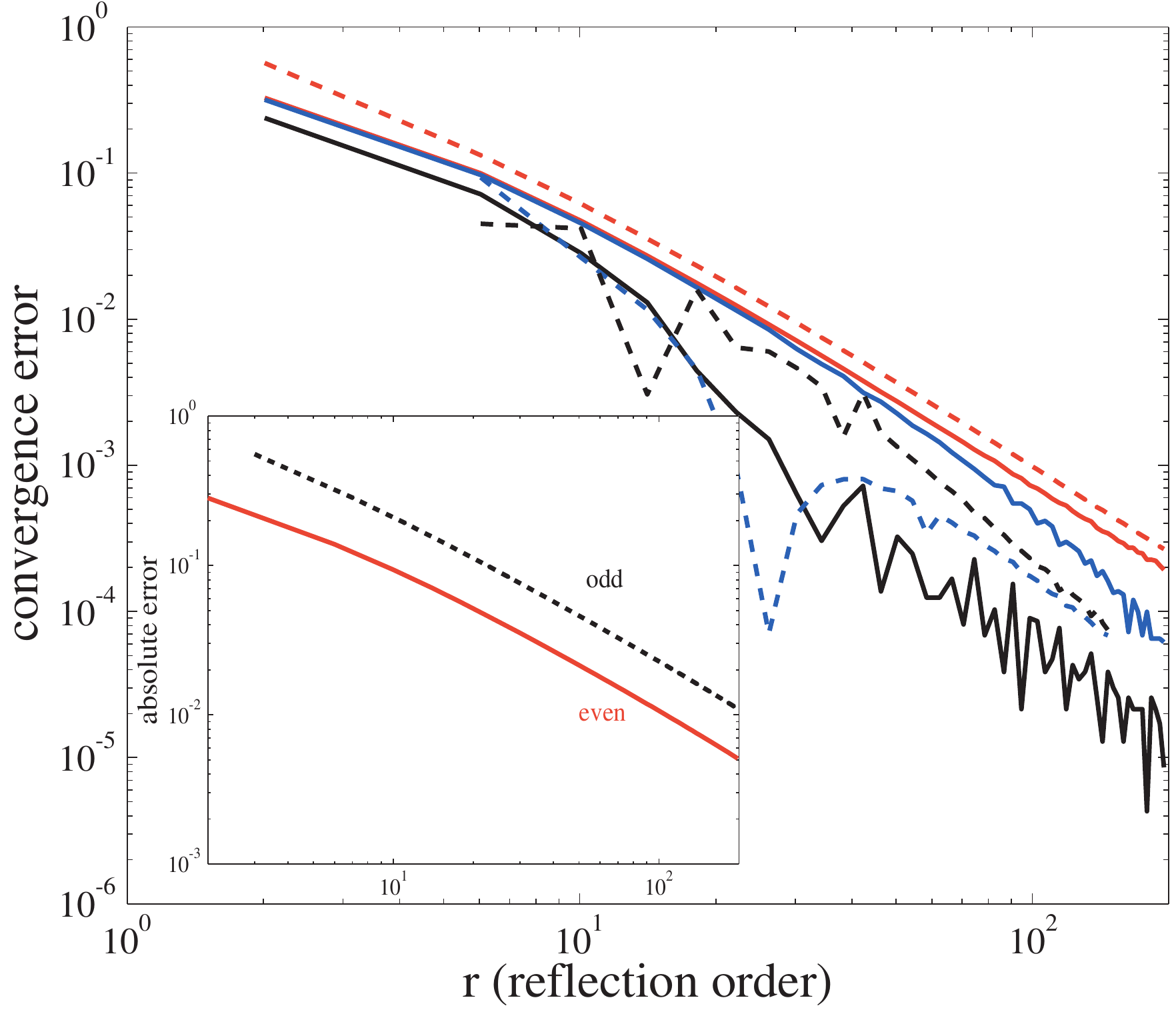}
\caption{(Color:) Convergence error
  $(\mathcal{E}_{r+1}-\mathcal{E}_r)/\mathcal{E}_r$ in the even
  (solid) and odd (dashed) path contributions vs. reflection order $r$
  for values of $h=0$ (red), $h=0.01$ (blue), and $h=0.1$ (black)
  (here, $\mathcal{E}_r$ means the energy computed up to order $r$).
  \emph{Inset:} Absolute error
  $(\mathcal{E}_{\text{exact}}-\mathcal{E}_r)/\mathcal{E}_{\text{exact}}$
  for both even (solid black) and odd (dashed red) contributions for
  $h=0$.}
\label{fig:conv}
\end{figure}

We repeat this calculation for increasing reflection order $r$ until
the total energy $\mathcal{E}_r$ converges to the desired accuracy.
From general considerations, one expects the error
$|\mathcal{E}-\mathcal{E}_r|$ for the energy from a finite $r$ to
decrease as $O(1/r)$.  In particular, the path lengths $\ell$ increase
proportional to $r$ (the radius in the extended lattice), and the
number of paths with a given length also increase proportional to $r$
(the circumference in the extended lattice), so the $\sum 1/\ell^3$
for a given $r$ goes as $O(1/r^2)$.  The error in the energy is the
sum over all paths of order $> r$, and this therefore goes as
$O(1/r)$.  This scaling is verified in \figref{conv}, which plots the
relative error $(\mathcal{E}_{r+1}-\mathcal{E}_r)/\mathcal{E}_r$
between the order-$r$ and the order-$(r+1)$ energy computations
for the particular case of $a=s=1$.  In general, if the energy
converged as $O(1/r^n)$ for some power $n$, one would expect this
difference to converge as $O(1/r^{n+1})$, and so we expect
\figref{conv} to asymptotically go as $O(1/r^2)$.  This is precisely
what is observed, for both even and odd paths, and for both $h=0$ and
$h>0$: all of the curves asymptotically approach straight lines (on
a log--log scale) with slope $-2$.

 An interesting though unfortunate result is that the odd-path energy
 requires larger $r$ in order to obtain the same accuracy as the
 even-path energy. Though this may not be obvious from looking at the
 convergence error, it is clear from the inset of \figref{conv}, where
 we plot the absolute error
 $(\mathcal{E}_{\text{exact}}-\mathcal{E}_r)/\mathcal{E}_{\text{exact}}$
 instead (at $h=0$). The constant offset observed in the absolute
 errors imply that, given a desired accuracy constrain on the even and
 odd energy calculations, one would have to compute roughly twice the
 number of odd paths in order to obtain equivalent accuracy.

\section{Concluding Remarks}

By comparing the ray-optics approximation with an exact
brute-force calculation, we have been able to study both the successes
and limitations of the ray-optics approximation. On the positive
side, the ray-optics approximation is capable of capturing surprising
behaviors that arise in closed geometries involving multiple bodies,
qualitatively matching phenomena identified in exact brute-force
calculations. In particular, the ray-optics approximation captures the
non-monotonic sidewall effects observed for metallic squares between parallel sidewalls,
generalized from the Casimir
piston geometry. This effect is clearly a manifestation of the
multi-body character of the interaction, since it does not arise in
simple two-body force laws such as PFA.  Ray optics
appears to be unique among the current simple approximations for
Casimir force in that it can capture such multi-body effects, even
though it cannot be quantitatively accurate in geometries with strong
curvature. On the negative side, diffractive effects set
in rather quickly when $h$ is increased from zero, marking the
agreement between ray-optics and the exact results only qualitative.

This makes the ray optics approximation a promising approach to
quickly search for unusual Casimir phenomena in complicated
geometries.  However, since it is an uncontrolled approximation in the
presence of strong curvature, any prediction by ray optics in such
circumstances must naturally be checked against more expensive exact
calculations.  There will undoubtedly be complex structures in the
future where ray optics fails qualitatively as well as quantitatively
For instance, ray optics has more difficulty with open
geometries---e.g., for two squares with only one sidewall, only PFA
paths are present.  On the other hand, the reach of the ray optics
technique seems in some sense to be larger than that of simpler
approximations such as PFA.

\section*{Acknowledgements}

This work was supported in part by the Nanoscale Science and
Engineering Center (NSEC) under NSF contract PHY-0117795 and by the
U.~S. Department of Energy (D.~O.~E.) under cooperative research
agreement \#DF-FC02-94ER40818 (RLJ). A.~Rodriguez was funded by a
D.~O.~E. Computational Science Graduate Fellowship under grant
DE--FG02-97ER25308. S.~Zaheer was funded by the MIT Undergraduate
Research Opportunities Program. We are also grateful to M.~Hertzberg
for sending his notes on the two-dimensional piston, and to
A.~Farjadpour and A.~McCauley for useful discussions.


\end{document}